\documentclass[twocolumn,preprintnumbers,amsmath,amssymb,superscriptaddress,prl]{revtex4}

\usepackage{graphicx}

\usepackage{amsmath}
\usepackage{amssymb}
\usepackage{dcolumn}
\usepackage{bm}
\usepackage{upgreek}
\graphicspath{{Figures/}}

\begin{document}
\preprint{}
\title{Strongly Enhanced Stimulated Brillouin Backscattering in an Electron-Positron Plasma}

\author{Matthew R. Edwards}
\email{mredward@princeton.edu}
\affiliation{Department of Mechanical and Aerospace Engineering, Princeton University, Princeton, New Jersey, 08544, USA}%
\author{Nathaniel J. Fisch}
\affiliation{Department of Astrophysical Sciences, Princeton University, Princeton, New Jersey, 08544, USA}%
\author{Julia M. Mikhailova}
\email{j.mikhailova@princeton.edu}
\affiliation{Department of Mechanical and Aerospace Engineering, Princeton University, Princeton, New Jersey, 08544, USA}%
\affiliation{Prokhorov General Physics Institute, Russian Academy of Sciences, 119991, Moscow, Russia}
\date{\today}

\begin{abstract}
Stimulated Brillouin backscattering of light is shown to be drastically enhanced in electron-positron plasmas, in contrast to the suppression of stimulated Raman scattering. A generalized theory of three-wave coupling between electromagnetic and plasma waves in two-species plasmas with arbitrary mass ratios, confirmed with a comprehensive set of particle-in-cell simulations, reveals violations of commonly-held assumptions about the behavior of electron-positron plasmas. Specifically, in the electron-positron limit three-wave parametric interaction between light and the plasma acoustic wave can occur, and the acoustic wave phase velocity differs from its usually assumed value.
\end{abstract}

\maketitle

Plasma interaction with electromagnetic fields is vital to the study of electron-positron plasmas, which appear in nature as a component of the early universe \cite{Ruffini2010} and in the vicinity of pulsars \cite{Goldreich1969,Ruderman1975}, quasars \cite{Wardle1998}, and black holes \cite{Blandford1977,Begelman1984}. Laboratory-created electron-positron plasmas have long been recognized as an exciting fundamental and technological opportunity for exploration of many astrophysical and anti-matter phenomena. Ongoing efforts to reproduce such plasmas in the laboratory \cite{Chen2015,Danielson2015} have recently culminated in a demonstration of a neutral and relatively dense ($10^{16}$ cm$^{-3}$) laser-produced electron-positron plasma \cite{Sarri2015}, yielding a path to laboratory observation of collective effects in pair plasmas and prompting examination of untested assumptions about the collective behavior of electrons and positrons. 

In the electron-positron limit, many standard plasma approximations break down due to the equal masses of the plasma components. Electron-positron plasmas are expected to exhibit unusual properties including enhanced solitary-wave phenomena, the absence of Faraday rotation, and strong nonlinear Landau damping \cite{Tsytovich1978}, as well as differences in the behavior of turbulence \cite{Helander2014}. In particular, although electromagnetic field interaction with density perturbations has been discussed \cite{Liu2011}, it is claimed that three-wave coupling (i.e. stimulated Raman and Brillouin scattering) entirely vanishes in an electron-positron plasma \cite{Tsytovich1978,Greaves1997,Liu2011,Tinakiche2012,Danielson2015} because the nonlinear current and charge density have a cubic dependence on charge \cite{Tsytovich1978}. However, since the transverse nonlinear current, which mediates backscattering, has a quartic dependence on charge, it does not cancel, and the above argument does not apply to the acoustic mode. An alternative picture for the suppression of stimulated Raman scattering is that the laser-driven ponderomotive force acts equally on electrons and positrons, so the net charge difference required for the formation of a Langmuir wave cannot develop, an argument which does not apply to stimulated Brillouin scattering because the acoustic mode does not require a net charge difference. 

Here we analytically and numerically study three-wave coupling in two-species plasmas where the components have comparable masses and equal temperatures, yielding a complete picture of stimulated Raman and Brillouin scattering. Differing from previous studies, our theory and numerical simulations predict significant stimulated Brillouin scattering in electron-positron plasmas, in contrast to the suppression of Raman scattering. Our use of fully kinetic particle-in-cell (PIC) simulations allows us to address the deficiencies of the two-fluid model in capturing the behavior of the electron-positron acoustic mode.

Plasma-based laser amplification by stimulated Raman or Brillouin backscattering of counter-propagating laser beams has been studied in detail \cite{Malkin1999,Ping2004,Lancia2010,Weber2013,Toroker2014,Malkin2007,Malkin2012,Lehmann2014,Malkin2014pop,Edwards2015} as a method for producing ultra-short pulses of extraordinarily high intensities by avoiding the compression gratings of chirped pulse amplification \cite{Strickland1985}. In Raman and Brillouin amplification, Langmuir and ion-acoustic waves, respectively, mediate the transfer of energy from a long pump laser pulse to a short, lower-frequency seed laser pulse. Ponderomotive forcing at the difference frequency of the two counter-propagating electromagnetic waves drives plasma fluctuations, which scatter pump photons into frequency-downshifted seed photons. When appropriately phase-matched, the fluctuations grow rapidly in time, producing massive amplification. Analysis of the governing equations leads to phase-matching conditions for the frequency ($\omega$) and wavevector ($\mathbf{k}$) of the pump, seed, and plasma waves, i.e.\ conservation of energy ($\omega_{\textrm{pump}}=\omega_{\textrm{seed}}+\omega_{\textrm{plasma}}$) and momentum ($\mathbf{k}_{\textrm{pump}}=\mathbf{k}_{\textrm{seed}}+\mathbf{k}_{\textrm{plasma}}$). With these relations, a counter-propagating geometry becomes a powerful tool for computationally or experimentally validating an analytically determined plasma dispersion relation $\omega(\mathbf{k})$, since resonant amplification will be observed at the $\mathbf{k}_{\textrm{seed}}$ which satisfies the phase-matching conditions.

Unlike those of an electron-ion plasma, the longitudinal modes of an electron-positron plasma are not separable by species. Instead of a Langmuir wave governed by the electron number density ($n_e$) and an acoustic wave driven primarily by the ion ($i$) dynamics, we have a plasma wave corresponding to charge density fluctuations ($\propto [n_e - n_i]$) and an acoustic wave with no electrostatic component corresponding to total density fluctuations ($n_e + n_i$) \cite{Zank1995}. Below, we derive a dispersion relation which connects the heavy-ion and electron-positron limits for arbitrary mass ratios in the range $0\le \beta=m_e/m_i\le 1$ ($m_s$ is mass of species $s$). Approaching the electron-positron limit by varying $\beta$, rather than varying the ion-positron ratio, provides an intuitive picture of the transition from the ion-acoustic wave ($\beta \to 0$) to the electron-positron acoustic wave ($\beta = 1$). Note that we will use $e$ and $i$ (electron and ion) to denote negatively and positively charged particles, though the results are applicable both to electron-positron plasmas ($\beta = 1$) and previously studied comparable-mass ion-ion plasmas, e.g. C$_{60}^-$/C$_{60}^+$ $(\beta \approx 1)$ \cite{Oohara2003,Oohara2005,Oohara2007}, Tl$^+$/I$^-$ $(\beta \approx 0.62)$ \cite{Schermann1978}, or Cs$^+$/UF$_6^-$ $(\beta \approx 0.38)$ \cite{Wexler1983}.


In a two-fluid treatment of the longitudinal modes, the one-dimensional species ($s = i,e$) continuity, species momentum, and Poisson equations formulated in terms of previously defined variables and species charge ($q_s$), velocity ($v_s$), partial pressure ($P_s$), and electric field ($E$)
\begin{align}
&\partial_t n_s +  \partial_x (n_s v_s) = 0 \\
&m_s n_s (\partial_t v_s + v_s \partial_x v_s) = -\partial_x P_s + q_s n_s E \\
&\partial_x E = 4 \pi e (n_i - n_e)
\end{align}
may be linearized and solved by assuming solutions of the form $e^{i(kx-\omega t)}$. We deal with pressure by setting $\partial_x P_s = \gamma_s T_s \partial_x n_s$, with $\gamma_s$ a correction factor for dropping the derivative of temperature ($T_s$) term from the derivative of the ideal gas law. Note that we only consider ions with one missing electron so that the species charges ($q_e = -q_i$) have the magnitude of a single electron charge (e), and for our initially neutral plasma $n_{e,0}=n_{i,0}$. For now, we will leave $\gamma_s$ unspecified, apart from observing that for a one-dimensional adiabatic process $\gamma_s = 3$, and for an isothermal process $\gamma_s=1$. The resultant coupled equations may be solved \cite{SM} for $\omega$ to yield:
\begin{equation}
\omega_{(L,A)}^2 = \frac{1}{2} \omega_{ek\beta}^2 \pm \frac{1}{2} \sqrt{\omega_{ek\beta}^4 - 4 k^2 C_e^2 \beta \left[ (1+\alpha)\omega_e^2 + \alpha k^2 C_e^2\right]}
\label{eqn:disp1}
\end{equation}
where $\omega_{ek\beta}^2 = (1+\beta)\omega_e^2 + (1+\beta\alpha)k^2 C_e^2$, $\omega_e^2 = 4 \pi n_{e,0} e^2 /m_e$, $C_s^2 = \gamma_s T_s / m_s$, $\alpha = \gamma_i T_i / \gamma_e T_e$, and $k = |\mathbf{k}|$. Langmuir waves ($L$) are given by the upper sign and acoustic waves ($A$) by the lower sign. 

For immobile ions ($\beta=0$), only the Langmuir wave solution exists, with $\omega_L^2=\omega_e^2+ C_e^2 k^2=\omega_e^2 + 3 T_e k^2/m_e$ ($\gamma_e=3$). To find the ion-acoustic dispersion relation for a heavy-ion plasma, we consider Eq.~\ref{eqn:disp1} in the limit $\beta \to 0$, $k^2 \to 0$, and $\alpha\to0$, since the ion-acoustic wave calculation is valid for $T_e \gg T_i$, yielding the standard $\omega_A^2=k^2 T_e/m_i$. Considering the electron-positron limit, we have $\beta=1$ and, in agreement with previous results \cite{Zank1995}, we find $\omega^2_{(L,A)}=\omega_e^2+k^2C_e^2\pm \omega_e^2$.

\begin{figure}
\centering
\includegraphics[width=\linewidth]{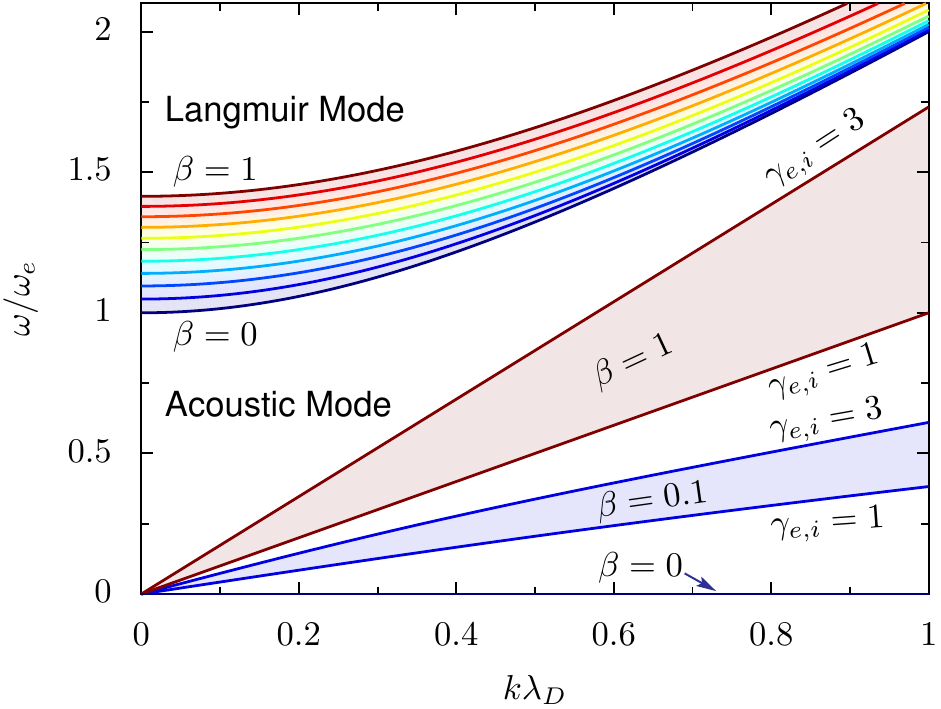}
\caption{Dispersion relations for the Langmuir and acoustic modes as $\beta = m_e/m_i$ is varied between 0 and 1 for $T_{e} = T_i$. $\lambda_D^2 =  T_e / 4 \pi e^2 n_e  =C_e^2/  \gamma_e^2 \omega_e^2$.}
\label{fig:dispRel}
\end{figure}

Due to the equivalent thermalization times of electrons and positrons, and our focus on $\beta > 0.1$, we will consider in detail only $\alpha = 1$. The resultant dispersion relation, valid for $0 \le \beta \le 1$, is, after some manipulation:
\begin{equation}
\omega_{(L,A)}^2 = \frac{1}{2} (1+\beta) \omega_{ek}^2 \pm \frac{1}{2} \sqrt{(1-\beta)^2 \omega_{ek}^4 + 4 \beta \omega_e^4}
\label{eqn:dispf}
\end{equation}
where $\omega_{ek}^2 = \omega_e^2 + k^2C_e^2$. By inspection, this equation still satisfies the electron-positron and immobile-ion limits. Equation~\ref{eqn:dispf} is plotted for $0 \le \beta \le 1$ in Fig.~\ref{fig:dispRel} at $T_{e,i} = 70$ eV and an electron number density $n_e = 10^{19}$ cm$^{-3}$. The Langmuir mode (upper curves) is characterized by the limits $\omega_L^2 \to (1+\beta)\omega_e^2$ as $k\to0$ and $\omega_L^2/k^2 \to 3T_e/m_e$ as $k \to \infty$, resulting from $\gamma_e = 3$, which is valid for all $\beta$, and $\gamma_i = 3$ in the regime $\beta \approx 1$ where positively-charged particles substantially affect the Langmuir wave. For the acoustic mode, the dispersion relations for $\gamma_s = 3$ and $\gamma_s = 1$ are both presented at $\beta=0,0.1,1$, with the region between the two values of $\gamma_s$ shaded, because the adiabatic assumption ($\gamma_s = 3$), which requires that the wave phase velocity is much greater than the species thermal velocity, is not valid for the acoustic mode. The similar phase and thermal velocities also result in Landau damping, so the acoustic wave is not easily observed in equal-mass plasmas \cite{Zank1995}. Though Eq.~\ref{eqn:dispf} suggests $\omega_{(L,A)}^2 / k^2 \to \gamma_e T_e/m_e$ as $k \to \infty$ for both modes, the different possible values of $\gamma_e$ means that the group velocities may differ in the large $k$ limit, in contrast to the usual assumption \cite{Zank1995,Verheest2006}. 

\begin{figure}
\centering
\includegraphics[width=\linewidth]{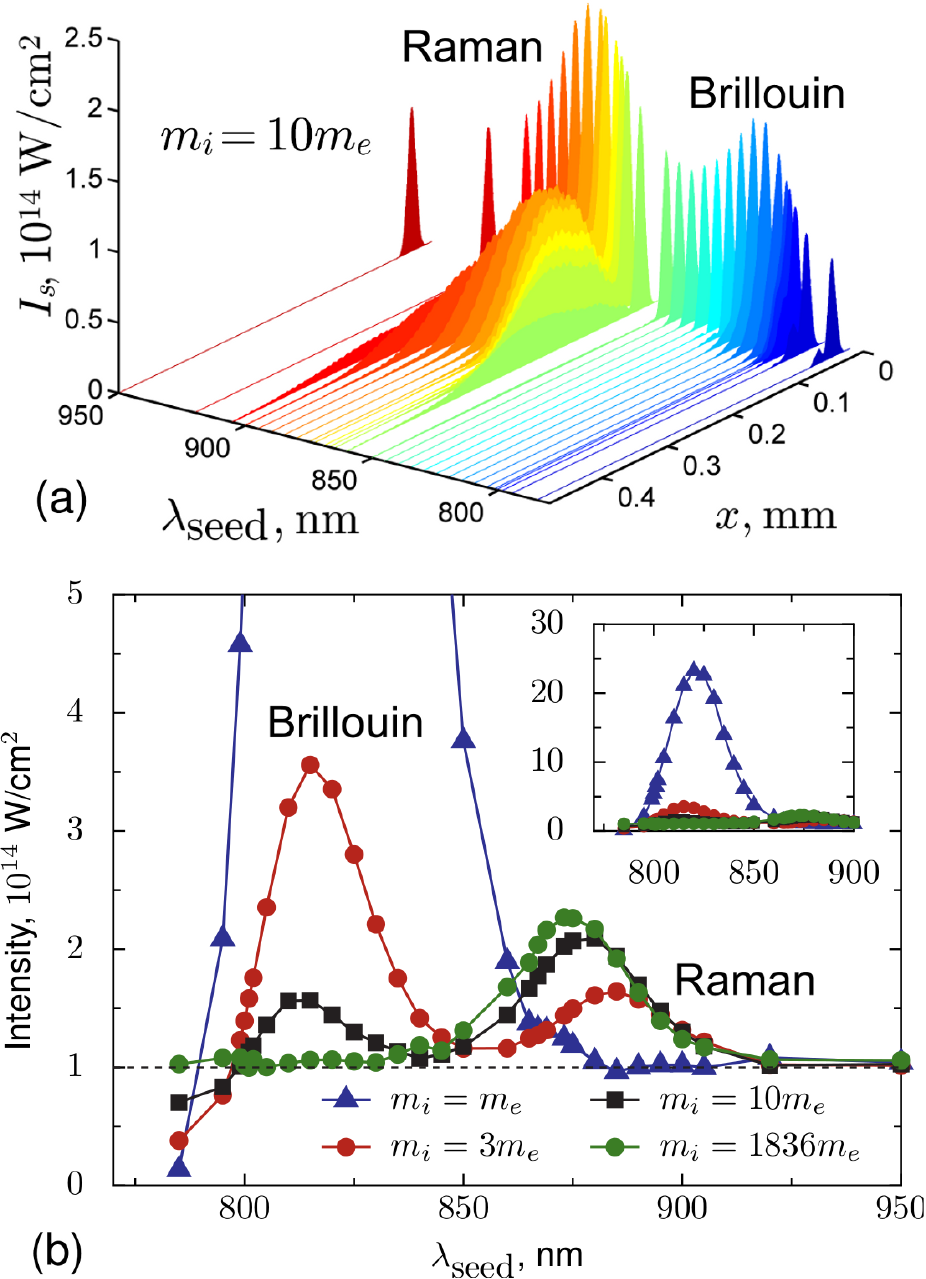}
\caption{(a) Amplified seed pulses of different wavelength ($\lambda_\textrm{seed}$) after passage through a 0.8 mm plasma with $n_{e,i} = 10^{19}$ cm$^{-3}$ and $m_i = 10 m_e$. Initial counter-propagating pump and seed intensities are $10^{14}$ W/cm$^2$ and $T_{e,i} = 70$ eV. (b) Maximum final intensity of an amplified seed at varied wavelength and ion mass $m_i$ with the same other parameters as (a). The dashed line indicates the initial seed intensity.}
\label{fig:Amp}
\end{figure}

Figure \ref{fig:Amp}a presents the effects of coupling between counter-propagating laser pulses in a 0.8 mm long, 70 eV plasma with $m_i = 10m_e$ ($\beta = 0.1$) and $n_e = 10^{19}$ cm$^{-3}$ ($0.0057n_c$) as found with fully-kinetic one-dimensional PIC simulations using the code EPOCH \cite{Arber2015}, showing the intensity envelope of amplified seed pulses of variable wavelength ($\lambda_{\textrm{seed}}$) after the interaction. Under these conditions, the lifetime of an electron-positron plasma is on the order of 10 $\mu$s, more than $10^8$ plasma wave periods \cite{Crannell1976,Gould1989,Iwamoto1993}. The pump (initial intensity $I_0 = 10^{14}$ W/cm$^2$) wavelength ($\lambda_\textrm{pump}$) is fixed at 800 nm as the seed ($I_0 = 10^{14}$ W/cm$^2$, intensity FWHM: 50 fs) wavelength is varied between 780 and 950 nm. The above parameters are also used in subsequent simulations, unless otherwise noted, with 80 cells/$\lambda_\textrm{pump}$ and 60 particles/cell. In Fig.~\ref{fig:Amp}a, two distinct resonances appear, near 875 nm (Raman) and 815 nm (Brillouin), giving the relationship between $\omega_{\textrm{plasma}}$ and $k_{\textrm{plasma}}$ at these plasma conditions; the different shapes of the intensity envelopes arise partially from the different damping behavior of the Langmuir and acoustic waves. The simulation parameters were chosen to be computationally tractable and allow comparison to previous results for Raman amplification at $\beta = 0$. 

To demonstrate how $\beta$ affects both the resonance wavelength and the instability growth rate, the final maximum intensity of the seed laser is plotted as a function of $\lambda_\textrm{seed}$ in Fig.~\ref{fig:Amp}b. Both the Raman and Brillouin resonances appear at longer seed wavelengths as $\beta \to 1$, indicating higher Langmuir and acoustic frequencies, and the Brillouin mode shows substantial enhancement.

We may consider in more detail the Raman (upper) solution to Eq.~\ref{eqn:dispf}. The heavy-ion ($\beta \to 0$) Langmuir wave neglects the ion contribution and takes $\gamma_e = 3$, which is valid where the electron thermal velocity is much lower than the Langmuir wave phase velocity. Since the thermal velocity of the ions is lower than that of the electrons, wherever $\gamma_e = 3$ is true, we can also take $\gamma_i = 3$. The Langmuir-wave phase velocity at phase-matched $k$ for the regime of interest ($k\lambda_D \approx 2 k_\textrm{pump}\lambda_D \approx 0.18$ at $T_{e,i}= 70$ eV, $n_e = 10^{19}$ cm$^{-3}$, and $\lambda_\textrm{pump} = 800$ nm) in an electron-positron plasma is higher than the particle thermal velocities, so the compression may be treated as adiabatic, justifying  $\gamma_{e,i}= 3$ for all $\beta$.

\begin{figure}
\centering
\includegraphics[width=\linewidth]{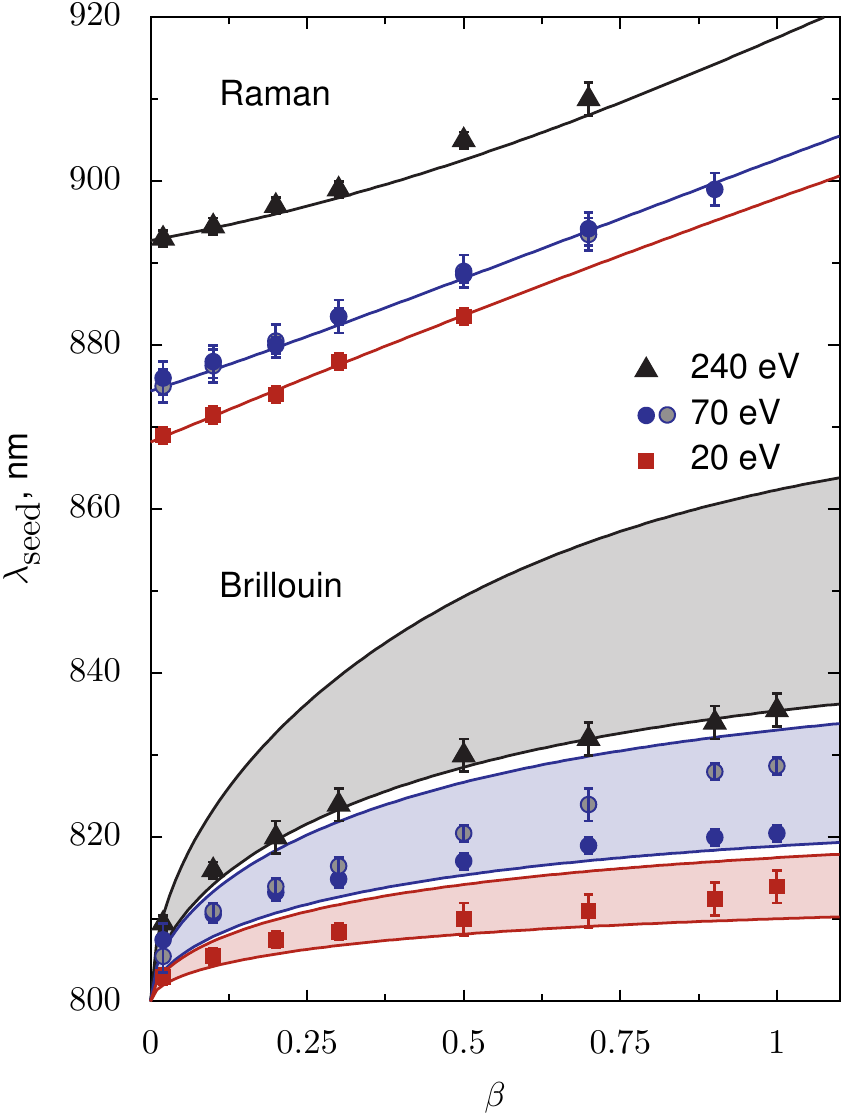}
\caption{Seed wavelengths ($\lambda_\textrm{seed}$) at which maximum Raman or Brillouin amplification is observed in PIC simulations (points) at varied plasma temperature ($T_{e,i} = 20$, $70$, $240$ eV) and ion mass ($\beta = m_e/m_i$), found by varying $\lambda_\textrm{seed}$ at fixed $\lambda_\textrm{pump} = 800$ nm. The solid lines are calculated from Eq.~\ref{eqn:dispf}. For the Brillouin mode, the upper and lower lines at each temperature correspond to $\gamma_{e,i} = 3$ and $\gamma_{e,i} = 1$, respectively. The 0.8 mm long plasma has a density $n_{e,i} = 10^{19}$ cm$^{-3}$ and $I_{\textrm{pump},0} = I_{\textrm{seed},0} = 10^{14}$ W/cm$^2$. The gray circles are calculated at a temperature of 70 eV and a higher pump intensity ($10^{15}$ W/cm$^2$) and overlap the lower intensity points for the Raman mode. The bars give uncertainty due to the finite size of changes in $\lambda_\textrm{seed}$ between simulations.}
\label{fig:SeedRaman}
\end{figure}

Figure \ref{fig:SeedRaman} shows the resonant seed wavelengths predicted analytically by Eq.~\ref{eqn:dispf} (solid lines) and determined from PIC simulations (points) by varying $\lambda_\textrm{seed}$ to find the value which results in the largest amplification. The theoretical predictions and simulation results agree for the Raman mode, suggesting that Eq.~\ref{eqn:dispf} captures the key dynamics of resonance for $0 \le \beta \le 1$. The pump intensity does not affect the Raman resonance wavelength in this regime, as shown by the overlap of the 70 eV results at two different pump strengths. Note that in Fig.~\ref{fig:SeedRaman} there are no simulation points at $\beta = 1$ for the Raman mode. This follows from the observation in Fig.~\ref{fig:Amp} that Raman mode amplification vanishes as $\beta \to 1$, as previously predicted \cite{Tsytovich1978}. 

Though we might hope that the acoustic mode, which in the heavy-ion limit is described by $\gamma_e = 1$ and $\gamma_i = 3$ \cite{Nicholson1983}, also approaches $\gamma_{e,i} =3$ in the electron-positron limit, the similarity of the acoustic wave phase velocity and particle thermal velocity means that wave propagation and thermalization are coupled, invalidating the adiabatic assumption. As Fig.~\ref{fig:SeedRaman} shows, the resonant $\lambda_\textrm{seed}$ for the Brillouin mode falls between the $\gamma_{e,i}=3$ and $\gamma_{e,i}=1$ solutions of the acoustic dispersion relation. Because the thermal and phase velocities are of the same order, thermalization of the velocity distributions occurs on the timescale of the wave period. Specifically, a non-negligible particle population travels multiple wavelengths in a single period and equilibrates the velocity distributions across the acoustic perturbations. Therefore, in the low-pump-intensity limit, the resonance condition for the electron-positron acoustic wave approaches the isothermal ($\gamma_s = 1$) rather than adiabatic ($\gamma_s = 3$) solution. This effect should be stronger (i.e. the resonance should be closer to $\gamma_s = 1$) at higher temperatures and lower intensities, in agreement with 70 eV Brillouin results of Fig.~\ref{fig:SeedRaman}. The counter-propagating geometry provides access to this difficult-to-study heavily-damped mode. A more precise analytic description of the acoustic resonance requires a kinetic approach, which lies beyond the scope of this paper.

\begin{figure}
\centering
\includegraphics[width=\linewidth]{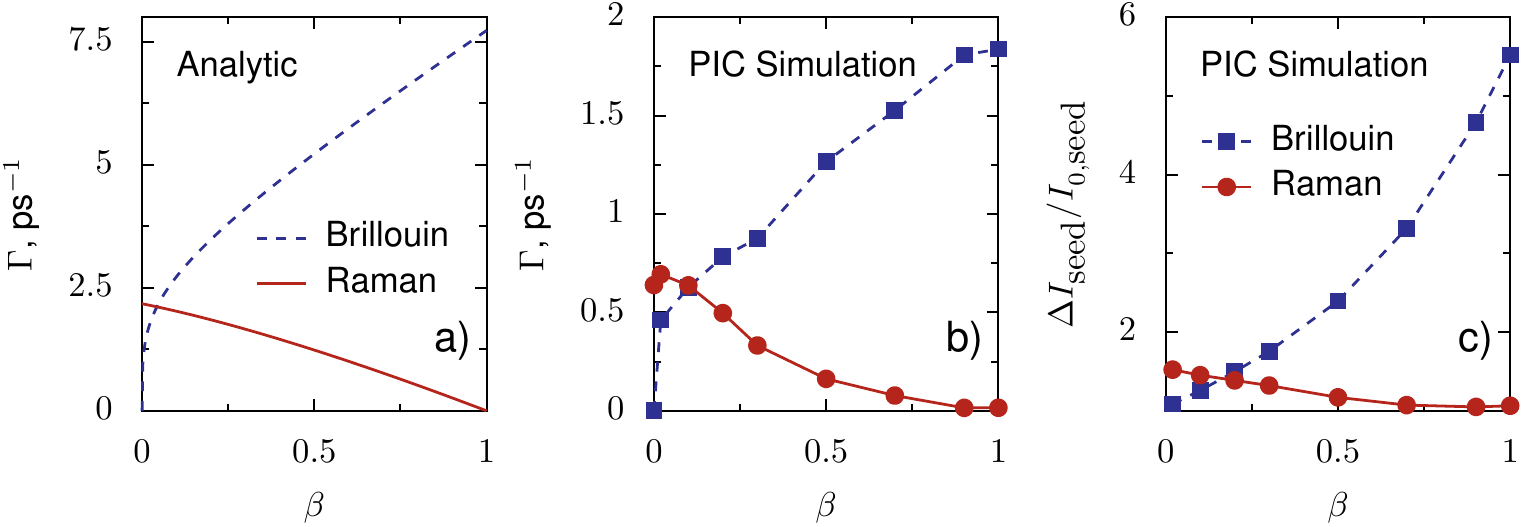}
\caption{Perturbation growth rate ($\Gamma$) for Raman and Brillouin scattering from the analytic dispersion relation (a) and found with PIC simulations below the saturation regime (b) at $T_e = T_i = 20$ eV and $n_e = 10^{19}$ cm$^{-3}$. In (b), the initial seed intensity is $10^{11}$ W/cm$^2$. (c) Simulated change in seed pulse intensity ($I_{\textrm{seed},0} = 10^{14}$ W/cm$^2$) after passage through 0.4 mm of plasma ($T_e = T_i = 70$ eV, $n_e = 10^{19}$ cm$^{-3}$).}
\label{fig:AmpRates}
\end{figure}

To analytically predict the amplification growth rate in arbitrary-ion-mass plasmas we require an equation for how a plasma perturbation mediates energy transfer between the pump and seed (vector potential $\mathbf{A}$) \cite{SM}
\begin{equation}
\left[ \partial_t^2 - c^2 \nabla^2 + (1+\beta)\omega_e^2\right] \mathbf{A}_\textrm{seed} = -\frac{4 \pi e^2}{m_e} \left[ \beta \tilde{n}_i + \tilde{n}_e\right] \mathbf{A}_\textrm{pump}
\label{eqn:maincoup1}
\end{equation}
where $\tilde{n}_s = n_s - n_{s,0}$ represents the density fluctuations, and a pair of equations describing how counter-propagating electromagnetic waves drive electron (Eq.~\ref{eqn:maincoup2}) and ion (Eq.~\ref{eqn:maincoup3}) plasma fluctuations \cite{SM}
\begin{align}
(\partial_t^2 - C_e^2 \nabla^2)\tilde{n}_e + \omega_e^2 (\tilde{n}_e - \tilde{n}_i) = F_{pe} \label{eqn:maincoup2} \\
(\partial_t^2 - \beta \alpha C_e^2 \nabla^2)\tilde{n}_i - \beta\omega_e^2 (\tilde{n}_e - \tilde{n}_i) = \beta^2 F_{pe}\label{eqn:maincoup3} 
\end{align}
where $F_{pe} = [n_{e,0} e^2/m_e^2 c^2] \nabla^2(\mathbf{A}_\textrm{pump}\cdot\mathbf{A}_\textrm{seed})$. Equations \ref{eqn:maincoup1}, \ref{eqn:maincoup2}, and \ref{eqn:maincoup3} may be linearized and combined to produce a single dispersion relation for the full system (see \cite{SM}). The substitution $\omega = \omega_{(L,A)} + \delta$ into the dispersion relation \cite{Kruer2003}, where $|\delta|\ll\omega_{(L,A)}$, gives an analytic formula for the instability growth rate of the seed field ($\Gamma = \textrm{Im}(\delta)$) \cite{SM}. For $\beta = 1$, $\Gamma_L = 0$ and $\Gamma_A = (\omega_e k e A_\textrm{pump} / 4 m_e c )\left[k C_e (\omega_\textrm{pump} - k C_e) /2\right]^{-1/2}$. The calculated growth rate for arbitrary $\beta$ is plotted in Fig.~\ref{fig:AmpRates}a, and the growth rate observed at corresponding conditions in PIC simulations is given in Fig.~\ref{fig:AmpRates}b. The growth rate observed in PIC simulations does not reach the maximum predicted analytically, partially due to the neglect of kinetic effects. Fig.~\ref{fig:AmpRates}c shows the change in seed intensity after passage through a 4 mm plasma, demonstrating that the general trend of amplification persists at intensities for which the seed reaches the saturation regime. In all three plots Brillouin scattering is strongly enhanced in the electron-positron plasma case. 

In summary, we have analyzed three-wave coupling in plasmas where the heavy-ion approximation does not hold. Because of their appearance in astrophysical phenomena and recent laboratory experiments, we emphasize electron-positron plasmas, though our results apply more generally. We show that the acoustic mode in an electron-positron plasma has a lower value of $\gamma_s$ than usually assumed in the literature. Most significantly, we find substantial stimulated Brillouin scattering in an electron-positron plasma, challenging the assumption that both Raman and Brillouin scattering are suppressed and suggesting scenarios where scattered radiation from electron-positron plasmas can be understood and used for diagnostics. 

\begin{acknowledgments}
This work is partially supported by the NSF under Grant No.~PHY 1506372 and the NNSA SSAA Program under Grant No.~DE274-FG52-08NA28553.  M.R.E. gratefully acknowledges the support of the NSF through a Graduate Research Fellowship. The computations reported in this paper were substantially performed at the TIGRESS high performance computer center at Princeton University. The EPOCH code was developed as part of the UK EPSRC 300 360 funded project EP/G054940/1.
\end{acknowledgments}

\onecolumngrid
%
%
%
%


\begin{center}
\textbf{\large Supplemental Material: Strongly Enhanced Stimulated Brillouin Backscattering in an Electron-Positron Plasma}
\end{center}
\setcounter{equation}{0}
\setcounter{figure}{0}
\setcounter{table}{0}
\setcounter{page}{1}

\section{The Longitudinal Dispersion Relation}

In a two-fluid treatment of the longitudinal modes, the one-dimensional species ($s = i,e$) continuity, species momentum, and Poisson equations, in terms of number density ($n_s$), mass ($m_s$), velocity ($v_s$), partial pressure ($P_s$), and electric field ($E$):
\begin{align}
&\partial_t n_s +  \partial_x (n_s v_s) = 0 \\
&m_s n_s (\partial_t v_s + v_s \partial_x v_s) = -\partial_x P_s + q_s n_s E \\
&\partial_x E = 4 \pi e (n_i - n_e)
\end{align}
may be linearized, with $n_s = n_0 + \tilde{n}_s$, $v_s = \tilde{v}_s$, and $E  = \tilde{E}$. We solve by assuming solutions of the form $e^{i(kx-\omega t)}$, setting $\partial_x P_s = \gamma_s T_s \partial_x \tilde{n}_s$, with $\gamma_s$ a correction factor for dropping the temperature derivative of the ideal gas law ($P_s = n_s T_s \to \partial_x P_s = \gamma_s T_s \partial_x n_s$) to give a relationship between only pressure and density. This produces the coupled equations:
\begin{equation}
\left[\begin{array}{c c} \omega^2 - k^2 C_e^2 - \omega_e^2 & \omega_e^2 \\ 
\omega_i^2 & \omega^2 - k^2 C_i^2 - \omega_i^2 \end{array} \right]
\left[\begin{array}{c} \tilde{n}_e \\ \tilde{n}_i \end{array} \right] = \mathbf{0}
\label{eqn:mat}
\end{equation}
where $C_s^2 = \gamma_s T_s / m_s$. This system has two eigenmodes, corresponding to solutions for Langmuir and acoustic waves. The dispersion relation is found by calculating the determinant of the matrix. Letting $\alpha = \gamma_i T_i / \gamma_e T_e$ so that $C_i = \beta \alpha C_e$, we have:
\begin{equation}
\left(\omega^2 - k^2 C_e^2 - \omega_e^2\right)\left(\omega^2 - \beta\alpha k^2 C_e^2 - \beta \omega_e^2\right) - \beta \omega_e^4= 0
\end{equation}
This is a quadratic equation in $\omega^2$ which can be solved to give:
\begin{equation}
\omega^2 = \frac{1}{2} \omega_{ek\beta}^2 \pm \frac{1}{2} \sqrt{\omega_{ek\beta}^4 - 4 k^2 C_e^2 \beta \left[ (1+\alpha)\omega_e^2 + \alpha k^2 C_e^2\right]}
\label{eqn:disp1}
\end{equation}
where $\omega_{ek\beta}^2 = (1+\beta)\omega_e^2 + (1+\beta\alpha)k^2 C_e^2$. Langmuir waves are given by the upper sign and acoustic waves by the lower sign.

\section{Limits of the Dispersion Relation}
To demonstrate that the above dispersion relation agrees with previous results we show below the evaluation of the dispersion relation in various limits. To find the heavy-ion Langmuir wave, we take $\beta = 0$. The above relation immediately simplifies to:
\begin{equation}
\omega^2 = \frac{1}{2} \omega_{ek}^2 \pm \frac{1}{2} \sqrt{\omega_{ek}^4}
\end{equation}
where $\omega_{ek}^2 = \omega_e^2 + C_e^2 k^2$. The lower operator gives $\omega^2 = 0$ as expected, because the ion-acoustic wave cannot exist with immobile ions. The upper operator yields
\begin{equation}
\omega^2 = \omega_e^2 + C_e^2 k^2 = \omega_e^2 + 3 \frac{T_e}{m_e} k^2
\end{equation}
which is the well-known Langmuir-wave dispersion relation ($\gamma_e = 3$) \cite{Nicholson1983S}. 

In the isothermal electron-positron limit, $\beta = 1$ and $\alpha = 1$, yielding:
\begin{align}
\omega^2 &= \omega_{ek}^2 \pm \frac{1}{2} \sqrt{4 \omega_{ek}^4 - 4k^2 C_e^2 \left[2 \omega_e^2 + k^2 C_e^2\right]}\nonumber \\
&= \omega_{ek}^2 \pm  \sqrt{ \omega_{e}^4 + 2\omega_e^2 k^2 C_e^2 + k^4 C_e^4  - 2\omega_e^2 k^2 C_e^2 - k^4 C_e^4}\nonumber \\
&= \omega_e^2 + C_e^2 k^2 \pm \omega_e^2
\end{align}
This is more obvious when we use the modified form of the dispersion relation (Eq.~5 in main text): 
\begin{equation}
\omega^2 = \frac{1}{2} (1+\beta) \omega_{ek}^2 \pm \frac{1}{2} \sqrt{(1-\beta)^2 \omega_{ek}^4 + 4 \beta \omega_e^4}
\end{equation}
which for $\beta = 1$ immediately becomes:
\begin{equation}
\omega^2 = \omega_e^2 + C_e^2 k^2 \pm \omega_e^2
\end{equation}
in agreement with previous work \cite{Zank1995S}.

To find the dispersion relation for the ion-acoustic wave in a heavy-ion plasma we consider Eq.~\ref{eqn:disp1} in the limit $\beta \to 0$, $k^2 \to 0$, and $\alpha \to 0$, since the standard ion-acoustic wave calculation assumes $T_e \gg T_i$. Formally, we will take $C_e^2 k^2 \ll \omega_e^2$, $\beta \ll 1$ and $\alpha \ll 1$. Expansion of the dispersion relation gives:
\begin{equation}
\omega^2 = \frac{1}{2} \left[(1+\beta)\omega_e^2 + (1+\beta \alpha)k^2 C_e^2\right] \pm \frac{1}{2} \sqrt{\left[(1+\beta)\omega_e^2 + (1+\beta \alpha)k^2 C_e^2\right]^2 - 4 k^2 C_e^2 \beta \left[ (1+\alpha)\omega_e^2 + \alpha k^2 C_e^2\right]}
\end{equation}
We will consider only the term under the radical, which may be written:
\begin{equation}
(1+\beta)^2 \omega_e^4 \left[1 + \frac{1}{(1+\beta)^2 \omega_e^4} \left(2(1+\beta +\beta \alpha + \beta^2 \alpha)k^2 C_e^2 \omega_e^2 + (1 + 2 \beta \alpha + \beta^2 \alpha^2) k^4C_e^4 -4 (\beta+\beta\alpha) k^2 C_e^2 \omega^2 -4 \beta \alpha k^4 C_e^4\right) \right]
\end{equation}
This simplifies to:
\begin{equation}
(1+\beta)^2 \omega_e^4 \left[1 + \frac{1}{(1+\beta)^2 \omega_e^4} \left(2(1-\beta -\beta \alpha + \beta^2 \alpha)k^2 C_e^2 \omega_e^2 + (1 - 2 \beta \alpha + \beta^2 \alpha^2) k^4C_e^4\right) \right]
\end{equation}
In this limit, all but the first term inside the radical are now much less than 1. We will therefore use the expansion $\sqrt{1+x} \approx 1+ x/2$ to rewrite the dispersion relation, noting that the highest order contained in the $x^2$ term of this expansion is of order $k^4 C_e^4 / \omega_e^4$, which will later be shown to be justification for keeping only the term linear in $x$:
\begin{equation}
\omega^2 = \frac{1}{2} \left[(1+\beta)\omega_e^2 + (1+\beta \alpha)k^2 C_e^2\right] \pm \left[\frac{1}{2}(1+\beta)\omega_e^2 + \frac{1}{2} \frac{(1-\beta)(1-\beta\alpha)}{(1+\beta)} k^2 C_e^2 + \frac{1}{4} \frac{(1-\beta\alpha)^2}{(1+\beta)}\frac{k^4C_e^4}{\omega_e^2}\right]
\end{equation}
It may be noted trivially that in this limit the Langmuir mode dispersion relation is $\omega^2 = \omega_e^2$. Considering the ion-acoustic mode, we take the lower operator, canceling the $(1+\beta)\omega_e^2$ terms and noting that for small $x$, $(1-x)/(1+x) \approx 1 - 2x +2x^2$ to find:
\begin{equation}
\omega^2 = \frac{1}{2}(1+\beta\alpha)k^2 C_e^2 - \frac{1}{2}(1-2\beta+2\beta^2)(1-\beta\alpha) k^2 C_e^2 + \frac{1}{4} \frac{(1-\beta\alpha)^2}{(1+\beta)}\frac{k^4C_e^4}{\omega_e^2}
\end{equation}
This simplifies to:
\begin{equation}
\omega^2 = \beta k^2 C_e^2 + \beta\alpha k^2 C_e^2 - \beta^2 k^2 C_e^2  - \beta^2 \alpha k^2 C_e^2  - \beta^3 \alpha k^2 C_e^2 + \frac{1}{4} \frac{(1-\beta\alpha)^2}{(1+\beta)}\frac{k^4C_e^4}{\omega_e^2}
\end{equation}

We consider $k^2 C_e^2 /\omega_e^2 \ll \beta$ which is permissible here because $\beta$ is small but finite (determined by the ion mass), whereas $k$ can in principle be made arbitrarily small. The first term in the above expression is therefore of highest order, and we may write:
\begin{equation}
\omega^2 = \beta k^2 C_e^2 = \frac{\gamma_e T_e}{m_i} k^2
\end{equation}
which is the ion-acoustic dispersion relation in the limit of small $k$ ($\gamma_e = 1$) \cite{Nicholson1983S}.

\section{Density Fluctuations}

\begin{figure}
\centering
\includegraphics[width=.4\linewidth]{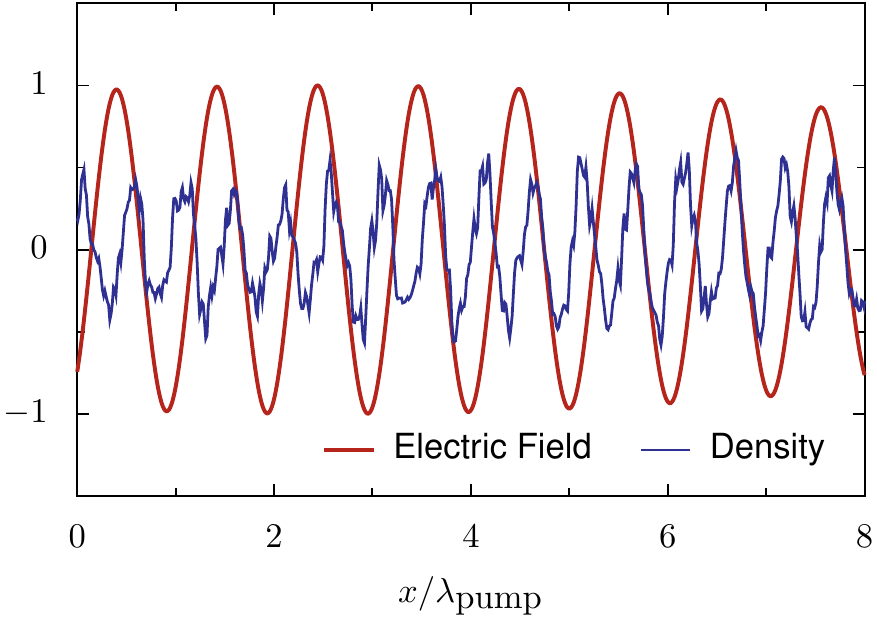}
\caption{Fluctuations in density and transverse electric field in an electron-positron plasma during intersection of counter-propagating laser pulses.}
\label{fig:DensityFluc}
\end{figure}

To check the interpretation of the Brillouin resonance as coupling to density fluctuations, the electron and positron density fluctuations found are plotted with the electric field in Fig. \ref{fig:DensityFluc}. It is clear from this image that the wavenumber of the density fluctuations is about twice that of the electric field, as is expected from the phase matching conditions. 

\section{Derivation of Growth Rate}
To derive an expression for the growth rate of the seed laser amplified by stimulated Raman or Brillouin backscattering we must find equations that describe both the effect of plasma perturbations on the transfer of energy from pump to seed and the creation of plasma perturbations from the interaction of pump and seed. We will extend the treatment of Kruer \cite{Kruer2003S} to cover an arbitrary-$\beta$ plasma. We may rewrite Ampere's law
\begin{equation}
\nabla \times \mathbf{B} = \frac{4\pi}{c} \mathbf{J} + \frac{1}{c} \partial_t \mathbf{E}
\end{equation}
in terms of potentials ($\mathbf{B} = \nabla \times \mathbf{A}$, $E = [-1/c] \partial_t \mathbf{A} - \nabla \phi$) to get:
\begin{equation}
\left[-\nabla^2  + \frac{1}{c^2} \partial_t^2 \right]\mathbf{A} = \frac{4\pi}{c} \mathbf{J} - \frac{1}{c} \partial_t \nabla \phi
\end{equation}
where we have chosen the Coulomb gauge to set $\nabla \cdot \mathbf{A} = 0$ and used $\nabla \times (\nabla \times \mathbf{A}) = \nabla(\nabla \cdot \mathbf{A}) - \nabla^2 \mathbf{A}$. Considering the transverse direction, we now have
\begin{equation}
\left[\frac{1}{c^2} \partial_t^2 - \nabla^2\right]\mathbf{A} = \frac{4\pi}{c} \mathbf{J}_\perp
\label{eqn:sm1}
\end{equation}
where $\mathbf{J}_\perp = en_i\mathbf{v}_i^\perp - en_e\mathbf{v}_e^\perp$. In the non-relativistic case we have:
\begin{equation}
\partial_t \mathbf{v}_s^\perp = \frac{q_s}{m_s}\mathbf{E}^\perp  = -\frac{q_s}{m_s c} \partial_t \mathbf{A}
\end{equation}
So that we may rewrite Eq. \ref{eqn:sm1} as:
\begin{equation}
\left[ \partial_t^2 - c^2 \nabla^2\right] \mathbf{A} = -4 \pi e^2 \left[ \frac{n_i}{m_i} + \frac{n_e}{m_e}\right] \mathbf{A} = -\frac{4 \pi e^2}{m_e} \left[ \beta n_i + n_e\right] \mathbf{A}
\end{equation}
We then substitute $\mathbf{A} = \mathbf{A}_\textrm{pump} + \mathbf{A}_\textrm{seed}$, noting that if $\omega_\textrm{pump} - \omega_\textrm{plasma} = \omega_\textrm{seed}$, the component of this equation at the frequency $\omega_\textrm{seed}$ becomes:
\begin{equation}
\left[ \partial_t^2 - c^2 \nabla^2 + (1+\beta)\omega_e^2\right] \mathbf{A}_\textrm{seed} = -\frac{4 \pi e^2}{m_e} \left[ \beta \tilde{n}_i + \tilde{n}_e\right] \mathbf{A}_\textrm{pump}
\end{equation}
where $\tilde{n}_s = n_s - n_{s,0}$. Assuming an exponential solution, this may be rewritten as:
\begin{equation}
\left[\omega^2 - c^2 k^2 -(1+\beta)\omega_e^2\right] \mathbf{A}_\textrm{seed}(k, \omega) = \frac{4 \pi e^2}{2 m_e} \left[n(k-k_\textrm{pump}, \omega-\omega_\textrm{pump}) + n(k+k_\textrm{pump}, \omega+\omega_\textrm{pump}) \right] \mathbf{A}_\textrm{pump}
\label{eqn:coup1}
\end{equation}
where $n = \beta \tilde{n}_i + \tilde{n}_e$.

Calculating the effect of the seed laser on the growth of ion and electron perturbations is more complex. We start with conservation of momentum:
\begin{equation}
\partial_t \mathbf{v}_s + \mathbf{v}_s \cdot \nabla \mathbf{v}_s = \frac{q_s}{m_s} \left[ \mathbf{E} + \mathbf{v}_s \times \mathbf{B}\right] - \frac{\nabla P_s}{n_s m_s}
\end{equation}
if we separate the velocity into transverse ($\mathbf{v}_s^\perp$) and longitudinal ($\mathbf{v}_s^\parallel$) components, where
\begin{equation}
\mathbf{v}_s = \mathbf{v}_s^\parallel + \mathbf{v}_s^\perp = \mathbf{v}_s^\parallel - \frac{q_s \mathbf{A}}{m_s c}
\end{equation}
The longitudinal momentum equation can be rewritten as:
\begin{equation}
\partial_t \mathbf{v}_s^\parallel = -\frac{q_s}{m_s} \nabla \phi - \frac{1}{2} \nabla \left( \mathbf{v}_s^\parallel - \frac{q_s \mathbf{A}}{m_s c} \right)^2 - \frac{\nabla P_s}{n_s m_s}
\end{equation}
Linearizing, with $\mathbf{v}_s^\parallel = \mathbf{\tilde{v}}_s^\parallel$, $n_s = n_0 + \tilde{n}_s$ ($n_{e,0} = n_{i,0}$), and $\mathbf{A} = \mathbf{A}_\textrm{pump} + \mathbf{A}_\textrm{seed}$, as well as taking the pressure derivative as $\nabla P_s = \gamma_s T_s \nabla n_s$, we arrive at:
\begin{equation}
\partial_t \mathbf{v}_s^\parallel = - \frac{q_s}{m_s} \nabla \tilde{\phi} - \frac{q_s^2}{m_s^2 c^2} \nabla \left( \mathbf{A}_\textrm{pump} \cdot \mathbf{A}_\textrm{seed} \right) - \frac{\gamma_s T_s}{n_0 m_s} \nabla \tilde{n}_s
\label{eqn:sm2}
\end{equation}
We now take the time derivative of the linearized continuity equation
\begin{equation}
\partial_t^2 \tilde{n}_s + n_0 \nabla \cdot \partial_t \mathbf{\tilde{v}}_s = 0
\end{equation}
And, noting that $\nabla \cdot \mathbf{v}_s^\perp = 0$, substitute in Eq.~\ref{eqn:sm2} to get:
\begin{equation}
\partial_t^2 \tilde{n}_s + n_0\left[\frac{-q_s}{m} \nabla^2 \tilde{\phi} - \frac{q_s^2}{m_s^2 c^2} \nabla^2 \left( \mathbf{A}_\textrm{pump} \cdot \mathbf{A}_\textrm{seed} \right) - \frac{C_s^2}{n_0} \nabla^2 \tilde{n}_s \right] = 0
\end{equation}
Application of the Poisson equation ($\nabla^2 \tilde{\phi} = 4 \pi e (\tilde{n}_e - \tilde{n}_i)$) allows the equations for $s = i,e$ to be written solely in terms of the density fluctuation magnitudes and the vector potentials:
\begin{align}
(\partial_t^2 - C_e^2 \nabla^2)\tilde{n}_e + \omega_e^2 (\tilde{n}_e - \tilde{n}_i) = \frac{n_0e^2}{m_e^2c^2} \nabla^2(\mathbf{A}_\textrm{pump}\cdot\mathbf{A}_\textrm{seed}) \\
(\partial_t^2 - \beta \alpha C_e^2 \nabla^2)\tilde{n}_e - \beta\omega_e^2 (\tilde{n}_e - \tilde{n}_i) = \beta^2 \frac{n_0e^2}{m_e^2c^2} \nabla^2(\mathbf{A}_\textrm{pump}\cdot\mathbf{A}_\textrm{seed}) 
\end{align}
These two equations describe how counter-propagating beams drive fluctuations in the electron and ion populations. 

To couple together the effect of the plasma fluctuations on the laser pulses with the effect of the lasers on the plasma fluctuations, we may rewrite the above equations in matrix form:
\begin{equation}
MN = \left[\begin{array}{c c} \omega^2 - k^2 C_e^2 - \omega_e^2 & \omega_e^2 \\ 
\beta \omega_e^2 & \omega^2 - \beta k^2 C_e^2 - \beta \omega_e^2 \end{array} \right]
\left[\begin{array}{c} n_e \\ n_i \end{array} \right] = \left[\begin{array}{c} -F_{pe} \\ -\beta^2 F_{pe} \end{array} \right] = -\left[\begin{array}{c}1 \\ \beta^2 \end{array} \right]F_{pe}
\label{eqn:coup2}
\end{equation}
where:
\begin{equation}
F_{pe} = \frac{n_0 e^2}{m_e^2 c^2} \nabla^2(\mathbf{A}_\textrm{pump}\cdot\mathbf{A}_\textrm{seed})
\end{equation}
Again assuming an exponential solution, the forcing term is:
\begin{equation}
F_{pe} = -\frac{n_0 e^2 k^2}{2 m_e^2 c^2} A_\textrm{pump} \left[A_\textrm{seed}(k-k_\textrm{pump}, \omega-\omega_\textrm{pump}) +A_\textrm{seed}(k+k_\textrm{pump}, \omega+\omega_\textrm{pump})\right]
\end{equation}
We may now substitute Eq.~\ref{eqn:coup1} to remove $A_\textrm{seed}$:
\begin{equation}
F_{pe} = -\frac{k^2 e^2 n_0}{2 m_e^2 c^2} A_\textrm{pump}^2 \frac{4 \pi e^2}{2 m_e} \left[ \frac{n(k, \omega)}{D(\omega+\omega_\textrm{pump}, k+k_\textrm{pump})} +  \frac{n(k, \omega)}{D(\omega-\omega_\textrm{pump}, k-k_\textrm{pump})}\right]
\label{eqn:coup3}
\end{equation}
where
\begin{equation}
D(\omega, k) = \omega^2 - k^2 c^2  - (1+\beta)\omega_e^2
\end{equation}
For amplification by backscattering we may neglect the contribution of the upshifted light. We would like to write Eq.~\ref{eqn:coup2} as an equation for $n$, so that it may be combined with Eq.~\ref{eqn:coup3}, which requires inverting $M$:
\begin{equation}
n = -\left[\begin{array}{c c} 1 & \beta \end{array} \right]M^{-1} \left[\begin{array}{c} 1 \\ \beta^2 \end{array} \right] F_{pe}
\end{equation}
With $v_{osc} = e A_\textrm{pump} / m_e c$ and $\omega_e^2 = 4\pi n_0 e^2 / m_e$, the two equations may be combined to give:
\begin{equation}
n(k, \omega) = \frac{\omega_e^2 k^2 v_{osc}^2}{4} \left[\begin{array}{c c} 1 & \beta \end{array} \right]M^{-1} \left[\begin{array}{c} 1 \\ \beta^2 \end{array} \right]  \frac{n(k,\omega)}{D(k-k_\textrm{pump}, \omega-\omega_\textrm{pump})}
\end{equation}
Writing the matrix $M$ as:
\begin{equation}
M =  \left[\begin{array}{c c} \omega^2 - k^2 C_e^2 - \omega_e^2 & \omega_e^2 \\ 
\beta \omega_e^2 & \omega^2 - \beta k^2 C_e^2 - \beta \omega_e^2 \end{array} \right] = 
\left[\begin{array}{c c}a &b \\ 
c &d \end{array} \right]
\end{equation}
This equation becomes:
\begin{equation}
\left[ (\omega-\omega_\textrm{pump})^2 - (k-k_\textrm{pump})^2 c^2  - (1+\beta)\omega_e^2 \right] (ad-bc) = \frac{\omega_e^2 k^2 v_{osc}^2}{4}\left( d - c \beta - b \beta^2 + a \beta^3\right)
\end{equation}
To find the growth rate we will substitute $\omega = \omega_{(L,A)} + \delta$ and determine the imaginary component of $\delta$, where $\omega_{(L,A)}$ is the Langmuir or acoustic resonance frequency. We will assume $|\delta| \ll \omega_{(L,A)}$. For clarity, we can divide the above equation into four components, with the first becoming:
\begin{align}
(\omega_{(L,A)} + \delta -\omega_\textrm{pump})^2 - (k-k_\textrm{pump})^2 c^2  - (1+\beta)\omega_e^2  = \delta^2& + 2 \delta(\omega_{(L,A)}-\omega_\textrm{pump})\:  + \nonumber \\ &\left[(\omega_{(L,A)}-\omega_\textrm{pump})^2 - (k-k_\textrm{pump})^2 c^2  - (1+\beta)\omega_e^2 \right]
\end{align}
In fulfilling the resonance condition, the bracketed terms are zero and we are left with $2 \delta(\omega_{(L,A)}-\omega_\textrm{pump}) + \delta^2$. The second component is evaluated as:
\begin{align}
ad - bc &= \left[(\omega_{(L,A)}+\delta)^2 - k^2 C_e^2 - \omega_e^2\right]\left[(\omega_{(L,A)}+\delta)^2 - \beta k^2 C_e^2 - \beta \omega_e^2\right] - \beta \omega_e^4 \nonumber \\
&= \left(2 \delta \omega_{(L,A)} +\delta^2\right)\left[2\omega_{(L,A)}^2 -(1+\beta) k^2 C_e^2 - (1+\beta) \omega_e^2\right] + \left(2 \delta\omega_{(L,A)} + \delta^2\right)^2
\end{align}
We will also simplify the fourth component:
\begin{equation}
\left( d - c \beta - b \beta^2 + a \beta^3\right) = \left(1+\beta^3\right)\left(\omega_{(L,A)}^2 + 2\delta \omega_{(L,A)} +\delta^2\right) - \beta\left(1+\beta^2\right) k^2 C_e^2 - \beta\left(1+\beta\right)^2 \omega_e^2
\end{equation}
Combining the simplified terms, we have:
\begin{align}
\delta^2 &\left[2 \left(\omega_{(L,A)} - \omega_\textrm{pump}\right) + \delta\right]\left[2 \omega_{(L,A)} + \delta \right]\left[2 \omega_{(L,A)}^2 - (1+\beta) k^2 C_e^2 - (1+\beta) \omega_e^2 + 2 \delta \omega_{(L,A)} + \delta^2 \right] = \nonumber \\
& \frac{\omega_e^2 k^2 v_{osc}^2}{4} \left[\left(1+\beta^3\right)\left(\omega_{(L,A)}^2 + 2\delta \omega_{(L,A)} +\delta^2\right) - \beta\left(1+\beta^2\right) k^2 C_e^2 - \beta\left(1+\beta\right)^2 \omega_e^2 \right]
\end{align}
Since the right hand side terms without explicit factors of $\delta$ are non-zero, $\omega_e^2 k^2 v_{osc}^2 $ must be of order $\delta^2$ for the equality to hold to order $\delta^2$, an assertion which we have confirmed for our parameters of interest. The lowest order terms are therefore of order $\delta^2$ and, dropping higher terms, i.e. everything with an explicit factor of $\delta^3$ from the left hand side, or $\delta$ from the right hand side, the equation becomes:
\begin{align}
\delta^2 &4 \left(\omega_{(L,A)} - \omega_\textrm{pump}\right) \omega_{(L,A)}\left[2 \omega_{(L,A)}^2 - (1+\beta) k^2 C_e^2 - (1+\beta) \omega_e^2  \right] = \nonumber \\
&\frac{\omega_e^2 k^2 v_{osc}^2}{4} \left[\left(1+\beta^3\right)\omega_{(L,A)}^2 - \beta\left(1+\beta^2\right) k^2 C_e^2 - \beta\left(1+\beta\right)^2 \omega_e^2 \right]
\end{align}
The growth rate $\Gamma$ is given by the imaginary component of $\delta$:
\begin{equation}
\Gamma = \frac{\omega_e k v_{osc}}{4} \left[ \frac{\left(1+\beta^3\right)\omega_{(L,A)}^2 - \beta\left(1+\beta^2\right) k^2 C_e^2 - \beta\left(1+\beta\right)^2 \omega_e^2}{\left(\omega_\textrm{pump} - \omega_{(L,A)}\right) \omega_{(L,A)}\left[2 \omega_{(L,A)}^2 - (1+\beta) k^2 C_e^2 - (1+\beta) \omega_e^2  \right]} \right]^{\frac{1}{2}}
\label{eqn:growth2}
\end{equation}
For the Langmuir wave ($\omega_L^2 = \omega_e^2 + C_e^2 k^2$) in the heavy ion ($\beta = 0$) limit, this simplifies to the expected Raman growth rate \cite{Kruer2003S}:
\begin{equation}
\Gamma = \frac{k v_{osc}}{4} \left[ \frac{\omega_e^2}{\omega_{ek} \left(\omega_\textrm{pump} - \omega_{ek}\right)} \right]^{\frac{1}{2}} 
\end{equation}
where $\omega_{ek}^2 = \omega_e^2 + C_e^2 k^2$. In the electron-positron ($\beta = 1$) limit, where the two relevant frequencies are $\omega_L^2 = 2\omega_e^2 + C_e^2 k^2$ and $\omega_A^2 = k^2 C_e^2$, the above equation gives $\Gamma = 0$ for the Langmuir wave and 
\begin{equation}
\Gamma = \frac{k v_{osc}}{4} \left[ \frac{2 \omega_e^2}{k C_e \left(\omega_\textrm{pump} -kC_e\right)} \right]^{\frac{1}{2}} 
\end{equation}
for the acoustic wave. Equation 48 is plotted for varied $\beta$ in Fig.~4a of the main text and compared to PIC simulations.


\end{document}